\documentclass{article}
\usepackage{ijcai25}

\usepackage{times}
\usepackage{soul}
\usepackage{url}
\usepackage[hidelinks]{hyperref}
\usepackage[utf8]{inputenc}
\usepackage[small]{caption}
\usepackage{graphicx}
\usepackage{amsmath}
\usepackage{amsthm}
\usepackage{booktabs}
\usepackage{algorithm}
\usepackage{algorithmic}
\usepackage{multirow}
\usepackage[switch]{lineno}
\usepackage{amssymb}

\title{GRAPE: Heterogeneous Graph Representation Learning for Genetic Perturbation with Coding and Non-Coding Biotype}

\author{
Changxi Chi$^{1,2*}$
\and
Jun Xia$^{1,2*}$\and
Jingbo Zhou$^{1,2}$\and
Jiabei Cheng$^{3}$\and
Chang Yu$^{2}$\And
Stan Z. Li$^{2}$\\
\affiliations
$^1$Zhejiang University, Hangzhou\\
$^2$AI Lab, Research Center for Industries of the Future, Westlake University\\
$^3$Department of Automation, School of Electronic Information and Electrical Engineering, Shanghai Jiao Tong University\\
\small{* Equal contribution}
\emails
\{chichangxi, xiajun\}@westlake.edu.cn
}

\begin{document}

\maketitle

\begin{abstract}
    Predicting genetic perturbations enables the identification of potentially crucial genes prior to wet-lab experiments, significantly improving overall experimental efficiency. Since genes are the foundation of cellular life, building gene regulatory networks (GRN) is essential to understand and predict the effects of genetic perturbations. However, current methods fail to fully leverage gene-related information, and solely rely on simple evaluation metrics to construct coarse-grained GRN. More importantly, they ignore functional differences between biotypes, limiting the ability to capture potential gene interactions. In this work, we leverage pre-trained large language model and DNA sequence model to extract features from gene descriptions and DNA sequence data, respectively, which serve as the initialization for gene representations. Additionally, we introduce gene biotype information for the first time in genetic perturbation, simulating the distinct roles of genes with different biotypes in regulating cellular processes, while capturing implicit gene relationships through graph structure learning (GSL). We propose \textbf{GRAPE}, a heterogeneous graph neural network (HGNN) that leverages gene representations initialized with features from descriptions and sequences, models the distinct roles of genes with different biotypes, and dynamically refines the GRN through GSL. The results on publicly available datasets show that our method achieves state-of-the-art performance. The code for reproducing the results can be seen at the anonymous link: \url{https://anonymous.4open.science/r/GRAPE-EB39}.
\end{abstract}

\section{Introduction}

\begin{figure}[htbp]
    \centering
    \includegraphics[width=0.4\textwidth]{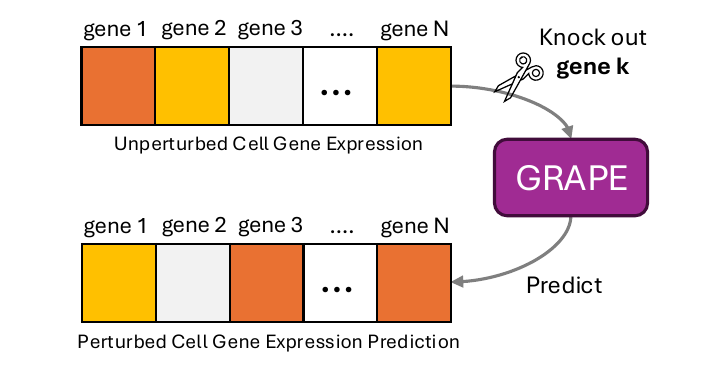}
    \caption{Schematic diagram of genetic perturbation.}
    \label{fig:Schematic}
\end{figure}

Single-cell genetic perturbation refers to the process of knocking out specific genes in single cells using CRISPR technology \cite{CRISPR_1,CRISPR_2}, thereby altering their normal physiological state or behavior. Despite significant advances in single-cell technology and high-throughput screening, it remains impractical to perform perturbation experiments on all genes within the vast genome of a cell. Accurate prediction of the effects of untested genetic perturbations helps reveal the specific roles of target genes in cellular biological processes and uncovers the regulatory relationships and network structures between genes. Using predictive models, potentially important genes can be identified before experiments, improving the efficiency of the utilization of experimental resources.\\
\indent Generative models have been widely applied in perturbation modeling, with existing approaches focusing on modeling the latent distribution of data to generate gene expression changes under various perturbations. SAMS-VAE \cite{sams-vae} enhances model interpretability by using sparse global latent variables to disentangle the specific effects of perturbations. GraphVCI \cite{graphVCI}, on the other hand, leverages graph convolutional networks to model gene regulatory networks (GRN) and predict gene expression under counterfactual perturbations. In contrast, GEARS \cite{GEARS} constructs a graph representation learning model based on gene pathways, ultimately producing prediction based on gene embeddings.\\
\indent However, these methods have several limitations. \textbf{1. Lack of Biological Information:} They do not fully leverage the rich biological information available in genes, such as detailed gene descriptions and DNA sequences, which provide valuable insights into the functional roles of genes. \textbf{2. Failure to Capture Implicit Gene Interactions:} Relying on simplistic evaluation metrics, these methods fail to capture the complex nature of gene regulation, where genes interact in diverse and multifaceted ways, hindering the accurate modeling and prediction of gene behavior across various biological contexts. \textbf{3. Ignoring Biotype-Specific Gene Functions:} Coding genes (c-genes) are transcribed into mRNA and translated into proteins, while non-coding genes (nc-genes) are transcribed into RNA but do not produce proteins. Biotype refers to the classification of genes into coding and non-coding categories based on their functional roles, with the functional dimensions of genes varying across biotypes \cite{nc_1,nc_3,nc_4,nc_5}. These methods overlook these differences, limiting their ability to model gene interactions effectively.\\
\indent To address these issues, we propose \textbf{GRAPE} (Heterogeneous \textbf{G}raph \textbf{R}epresentation Le\textbf{A}rning for Genetic \textbf{PE}rturbation) to predict genetic perturbation, which utilizes pre-trained large language model and DNA sequence model for gene feature initialization. Furthermore, \textbf{GRAPE} constructs a GRN as heterogeneous graph to capture the distinct roles of different biotypes and automatically learn the interactions between genes via graph structure learning. Experiments on publicly available datasets show that \textbf{GRAPE} outperforms existing methods.\\
\indent The main contributions of our work are as follows:
\begin{itemize}
\item \textbf{Perspective:} We introduce \textbf{GRAPE}, the first model to incorporate cellular gene biotype information (coding and non-coding genes), to model gene regulatory networks as heterogeneous graphs. This approach captures the distinct functional roles of genes from different biotypes through heterogeneous graph neural networks.
\item \textbf{Algorithm:} We initialize gene representations by leveraging pre-trained models \cite{bert,hyenaDNA} to extract multi-modal features of genes (textual descriptions and DNA sequences), which are then semantically aligned and fused to generate richer and more comprehensive representations. Besides, in constructing the gene interaction network, we propose a novel graph structure learning (GSL) method to capture implicit gene relationships.
\item \textbf{Experiments:} We demonstrate the superiority of \textbf{GRAPE} over existing methods on publicly datasets, using a range of comprehensive evaluation metrics.
\end{itemize}

\section{Related Work}
\subsection{Graph Structure Learning (GSL) and Heterogeneoous Graph Neural Network (HGNN)}
GSL optimizes the graph's structure to enhance information flow, improve node representations.\cite{GSL_1} jointly learn the graph structure and GCN parameters; SLAPS (\cite{GSL_2}) uses self-supervision to better infer graph structures for GNNs; HGSL(\cite{GSL_3}) generating relation subgraphs based on feature similarity, propagation, and semantics. HGNNs learn relationships between different node types and generate node representations. HetGNN \cite{HetGNN} uses a random walk strategy and two aggregation modules to generate embeddings; HAN \cite{HAN} uses hierarchical attention mechanisms at both node and semantic levels; SeHGNN \cite{SeHGNN} reduces complexity by using a lightweight mean aggregator and avoids excessive attention mechanisms. Despite their individual advantages, there remains room for improvement in capturing more complex and implicit relationships within heterogeneous graphs.

\subsection{Genetic Perturbation Model}
Gene perturbation is a essential direction in single-cell perturbation research. The Variational Autoencoder (VAE) framework and Graph Neural Networks (GNNs) have emerged as the dominant approaches in this filed of study. scGen \cite{scGen} is the first to use deep learning methods to model single-cell perturbations, utilizing variational autoencoders to capture cellular responses to perturbations in single-cell gene expression; GEARS \cite{GEARS} introduces ontology modality to model gene-gene interaction relationships; CellOracle \cite{CellOracle} leverages gene-regulatory networks from single-cell multi-omics data to simulate transcription factor perturbations; SAMS-VAE \cite{sams-vae} incorporates sparse global latent variables to enhance model interpretability; GraphVCI \cite{graphVCI} propose a novel graph-based Bayesian causal inference framework to predict gene expression responses. These methods have several limitations. They fail to fully leverage available biological information, which provide valuable insights into gene functions. Moreover, relying on simplistic evaluation metrics, they cannot capture the complex nature of gene regulation, hindering accurate predictions. Additionally, overlooking the functional differences between genes of different biotypes hinders their ability to effectively model gene interactions.

\section{Preliminaries}
In this section, we expound some basic concepts and give some definitions of our model.\\
\textbf{Definition 1. Heterogeneous Graph} \hspace{0.3cm} We introduce an undirected heterogeneous graph $G=(N,R,X)$, where $N$ is node set, along with node type mapping function $ \varphi: N \rightarrow T=\{ \text{c-gene}, \text{nc-gene} \}$, $R$ is edge set, along with edge type mapping function $ \phi: R \rightarrow E=\{ r_{cc}, r_{cn}, r_{nn} \}$, $X$ is node feature matrix.\\ 
\textbf{Definition 2. Model Input and Output} \hspace{0.3cm} The model takes as input gene textual descriptions and DNA sequence information, along with the perturbation conditions. As illustrated in Fig.\ref{fig:Schematic}, the task of our model is to predict the change in single-cell expression levels under a given perturbation condition (e.g., knockdown of gene $k$).


\begin{figure*}[htbp]
    \centering
    \includegraphics[width=0.84\textwidth]{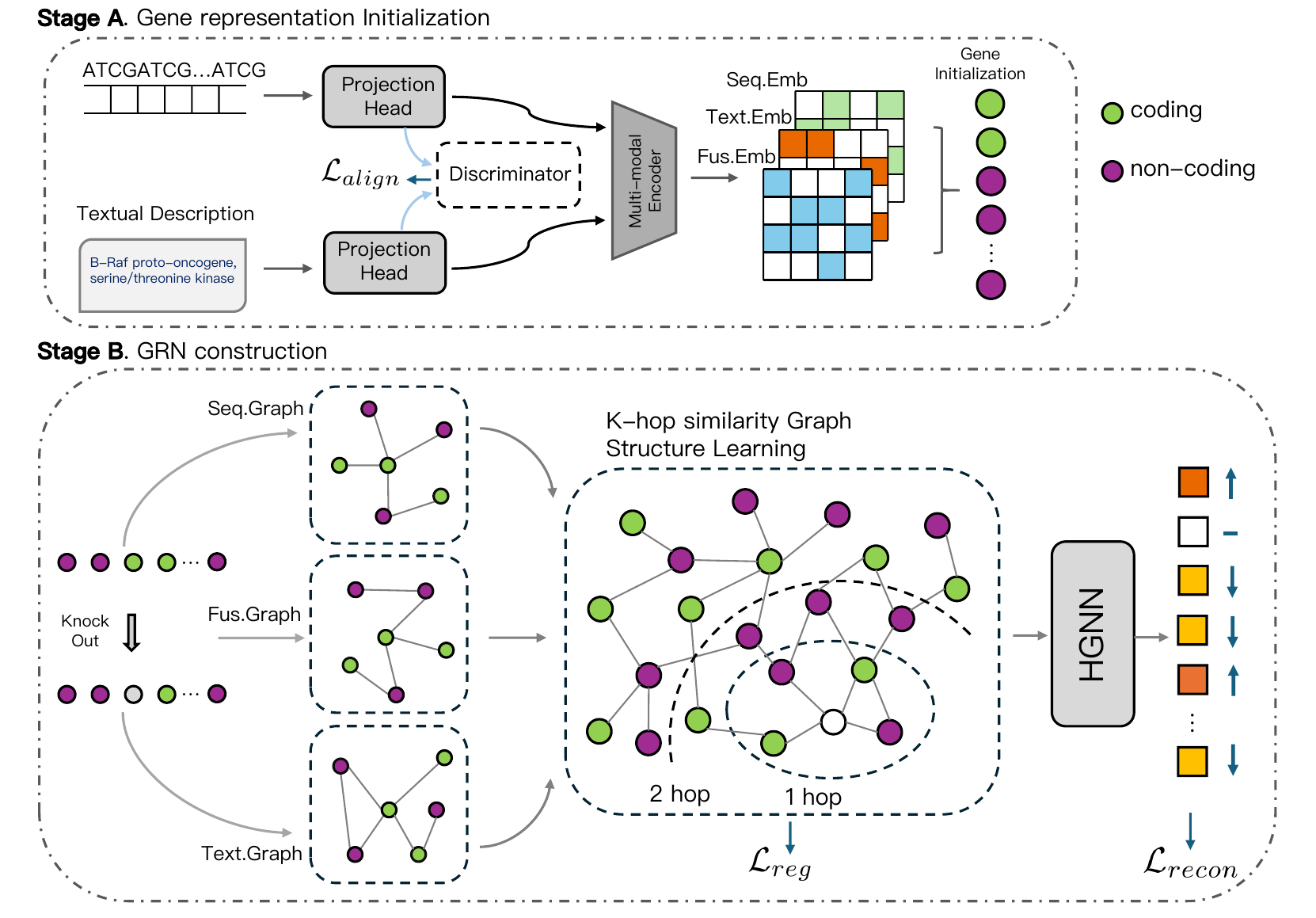}
    \caption{Overview of the \textbf{GRAPE}.}
    \label{fig:flowchart}
\end{figure*}

\section{Method}
In this section, we introduce the proposed model GRAPE. Figure.\ref{fig:flowchart}.A illustrates the workflow for initializing gene features using sequence and textual descriptions by DNA sequence model \cite{hyenaDNA} nad LLM \cite{bert}, respectively. Figure.\ref{fig:flowchart}.B presents the methodology for constructing gene interaction networks and performing heterogeneous graph representation learning.

\subsection{Gene Representation Initialization}
Existing methods often initialize gene representations based on gene indices, which lack biological prior information and interpretability, potentially increasing the cost of model convergence. Therefore, we propose multi-modal biological prior information for representation initialization, enabling a more effective capture of the intrinsic gene characteristics. Building upon the pre-trained models \cite{hyenaDNA,bert}, we align gene sequences with their textual descriptions to effectively integrate and capture the shared semantic information between them.\\
\indent Formally, we extract DNA sequence features using \cite{hyenaDNA} and textual description features using \cite{bert}, resulting in the multi-modal features matrix $X_m =[x_m^1,x_m^2,...,x_m^N] \in R^{N \times d_m}$ , where $m \in \{\tt{seq}, \tt{text}\}$ denotes either the sequence or the textual description modality, with a dimensionality of $d_m$. Our goal is to ensure that the semantic integrity of each modality is preserved while achieving a unified feature representation $Z_{m}=f(X_{m})=[z_{m}^{1},z_{m}^{2},...,z_{m}^{N}]^{T}\in {R}^{N \times q_m}$ after the alignment, where $f$ represents the transformation process.\\
\indent Specifically, we aim to align the features extracted from different modalities for the same gene, while ensuring that cross-modal features from different genes are pushed apart. We achieve this using contrastive learning \cite{contrastive_learning}, and the corresponding formula is as follows:
\begin{equation}
    \mathcal{L}_{align} = -\sum_{i=1}^{N}\frac{exp^{(D(z_{S}^{i},z_{D}^{i})/ \mathcal{T})}}{\sum_{j=1}^{N}exp^{ (D(z_{S}^{i},z_{D}^{j})/ / \mathcal{T})}}
    \label{eq:align}
\end{equation}
where $D$ is the discriminator function that estimates the cosine similarity between the features after transformation. This indirect contrastive learning approach helps mitigate the disruption of the original semantics of each modality during alignment. After modality alignment, we obtain the multi-modal fusion features through a multi-layer perceptron (MLP) $\Psi$:
\begin{equation}
    Z_{\tt{fus}}=\Psi(Z_{\tt{seq}}||Z_{\tt{text}})
    \label{eq:fusion}
\end{equation}

\subsection{Gene Regulation Network Construction}
Based on the multi-modal features previously extracted, we can construct a graph for each modality $m$, where $m\in M =\{\tt{seq},\tt{text},\tt{fus}\}$, according to the similarity of the characteristics between genes, thus revealing gene associations from different perspectives. Specifically, each gene is treated as a node in the graph, with multiple modality-specific representations. The edge between node $z_{m}^{i}$ and node $z_{m}^{j}$ is obtained by:
\begin{equation}
    G_{m}[i,j]=\begin{cases}
                \mathbb{S}_{m}(z_{m}^{i},z_{m}^{j}), & \text{if } \mathbb{S}_{m}(z_{m}^{i},z_{m}^{j}) \ge \epsilon^{m}\\
                0, & \text{otherwise}
                \end{cases}
    \label{eq:multi-view graph construction}
\end{equation}
where $\epsilon^{m}$ is used to control the sparsity of the graph corresponding to the $m$ modality. $\mathbb{S}_{m}$ is a multi-head similarity module that computes multiple similarity scores using different heads with independent parameters. The definition is as follow:
\begin{equation}
    \mathbb{S}_{m}(z_{m}^{i},z_{m}^{j})=\frac{1}{T}\sum_{t=1}^{T}cos(z_{m}^{i} W_{m}^{t}, 
    z_{m}^{j} W_{m}^{t})
    \label{eq:K-head similarity}
\end{equation}
where $W_{m}^{t}$ denotes the learnable parameters of the $t$-th similarity head in the $m$ modality. After the above graph construction steps and edge filtering, we obtain the gene interaction network structures $G_{m}$ from the perspective of each modality. To aggregate the Gene Regulatory Networks (GRNs) $G_{agg}$ from different modalities, we can express the aggregation as the mean across all modalities:
\begin{equation}
    G_{agg}=\frac{1}{|M|}\sum_{m \in M}G_{m}
\end{equation}
\indent However, graphs constructed using simple similarity-based methods tend to capture only coarse-grained neighbor relationships, making it challenging to comprehensively model the complex and latent interactions between nodes. In real-world scenarios, genes often exhibit latent interactions across multiple scales, necessitating a more flexible and holistic mechanism to model their associations. To address this, we introduce the $K$-hop GCN aggregation mechanism, which aggregates the results from different $K$-hop convolutional layers to integrate multi-scale structural information, enabling a more effective representation of the intricate gene interactions. In practice, we aggregate the node relationships across different hop distances as follows:
\begin{equation}
    H^{(l+1)}=\operatorname{GCN}(G_{agg},H^{(l)})
    \label{eq:multi-GCN}
\end{equation}
\begin{equation}
    \mathbb{G}^{k}[i,j]=\mathbb{S}^{k}(H^{k}_{i} , H^{k}_{j})
\end{equation}
when $l=0$, the feature matrix $H^{(l)}$ is typically initialized as the initial node fusion representations $Z_{\tt{fus}}$ (Eq.~(\ref{eq:fusion})), with $\operatorname{GCN}$ serving as the backbone for feature aggregation \cite{GCN}. \\
\indent Then the $K$-hop structure is defined as:
\begin{equation}
    \mathbb{G}=\Phi(\mathbb{G}^{1},\mathbb{G}^{2},\dots,\mathbb{G}^{k})
    \label{eq:final}
\end{equation}
\begin{equation}
    \mathbb{E}=\{(i,j)|\mathbb{G}(i,j)\ge \epsilon\}
    \label{eq:final_edge}
\end{equation}
where $\Phi$ denotes a channel attention layer. The edge set $\mathbb{E}$ are selected to capture the most significant relationships between nodes. To maintain the structural consistency between the original graph $G_{agg}$ and the $k$-hop graph $\mathbb{G}$, the regularization (Eq.~(\ref{eq:reg})) is proposed, ensuring that the model does not deviate from the local structural information of the original graph while learning higher-order graph structures.
\begin{equation}
    \mathcal{L}_{reg}=||\mathbb{G}-G_{agg}||_2
    \label{eq:reg}
\end{equation}

\subsection{Heterogeneous Graph Representation Learning}
As mentioned earlier, c-genes and nc-genes regulate the cellular state at different level \cite{nc_1,nc_3,nc_4,nc_5}. Therefore, it is necessary to construct a gene interaction network using a heterogeneous graph based on the gene biotype. To simplify the problem, we regard the heterogeneous graph as an undirected graph.\\
\indent Based on the graph construction before, we map the nodes using $ \varphi$ and the edges using $\phi$ as defined in Definition.1. Given the different types of associations between genes, we design a multi-head heterogeneous graph attention network (HGAT) to aggregate these features:
\begin{equation}
    P^{(l+1)}=\frac{1}{|E|}\sum_{r \in E}
    (\frac{1}{H}\sum_{h=1}^{H} \operatorname{GAT}_{r}^{h}
    ( \{ e|e \in \mathbb{E} \land \phi(e)=r \} ,P^{l}))
    \label{eq:HGAT}
\end{equation}
where $H$ represents number of head, and $P^{(0)}=Z_{\tt{fus}}$ is initialized as the fusion gene embeddings with perturbation information, which will be elaborated on in the following section. The $\operatorname{GAT}$ used for feature aggregation can be found in \cite{GAT,GAT2}.This framework enables the model to actively learn the weights of different types of edges by leveraging the attention mechanism. In the context of a heterogeneous graph, where various edge types represent distinct gene interactions, the model can assign independent parameters to each type of relationship. By doing so, it allows the model to prioritize the most relevant interactions for each specific task, rather than treating all the edges equally.

\begin{table*}[]
\centering
\resizebox{\textwidth}{!}{
\begin{tabular}{ccccccccc}
\toprule
\multirow{2}{*}{}                & \multirow{2}{*}{}       & \multirow{2}{*}{\textbf{RMSE($\downarrow$)}} & \multicolumn{3}{c}{\textbf{DAC($\uparrow$)}}               & \multicolumn{3}{c}{\textbf{PCC\_delta($\uparrow$)}}                   \\  
                                        \cmidrule(lr){4-6} \cmidrule(lr){7-9} 
                                 &                         &                                & \textbf{All}     & \textbf{DE20}   & \textbf{DE40}   & \textbf{All}     & \textbf{DE20}   & \textbf{DE40}   \\ \midrule
\multirow{4}{*}{\textbf{Adamson}} & \textbf{{GRAPE (Ours)}} & $\textbf{0.0754}\pm0.0006$         & $\textbf{0.6948}\pm0.0031$  & $\textbf{0.8863}\pm0.0029$  &  $\textbf{0.9363}\pm0.0023$ & $\textbf{0.7461}\pm0.0070$  & $\textbf{0.7828}\pm0.0124$ & $\textbf{0.8010}\pm0.0110$  \\
                                & GEARS                   & $0.0812\pm0.0008$        & $0.6557\pm0.0024$         & 
                                $0.7881\pm0.0038 $        & $0.7876\pm0.0027$        & $0.5514 \pm0.0064$        & $0.6109\pm0.0059$         & $0.6429\pm0.0084$         \\
                                & GraphVCI                & $0.4031\pm0.0064$        & $0.5315\pm0.0184 $        & 
                                $0.6114 \pm 0.0233$       & $0.4807\pm0.0112$        & $0.1647\pm0.0201 $        & 
                                $0.4138\pm 0.0306$        & $0.4405\pm0.0291$         \\
                                & sams-VAE                & $0.6511\pm0.0113$        & $0.4241\pm0.0218$         & $0.6659\pm0.0351$         & $0.5625\pm0.0377$        & $0.1441\pm0.0195$         & $0.4155\pm0.0432$         & $0.4313\pm0.0319$         \\ \midrule
\multirow{4}{*}{\textbf{Norman}}  & \textbf{{GRAPE (Ours)}} & $\textbf{0.0482}\pm0.0003$           & $\textbf{0.7583}\pm0.0012$ & $\textbf{0.7352}\pm0.0014$ & $\textbf{0.7556}\pm0.0011$ & $\textbf{0.4742}\pm0.0052$ & $\textbf{0.5452}\pm0.0152$ & $\textbf{0.5581}\pm0.0107$  \\
                                & GEARS                   & $0.0702\pm0.0007$         & $0.5201\pm0.0050$         & 
                                $0.6319\pm0.0057$         & $0.6361\pm0.0047$         & $0.4277\pm0.0036$         & 
                                $0.5012\pm0.0044$         & $0.5417\pm0.0043$         \\
                                & GraphVCI                & $0.5572\pm0.0187$         & $0.2795\pm0.0127$         & 
                                $0.3395\pm0.0183$         & $0.3664\pm0.0174$         & $0.1202\pm0.0147$         & 
                                $0.3205\pm0.0340$         & $0.2423\pm0.0286$         \\
                                & sams-VAE                & $0.5025\pm0.0170$         & $0.3742\pm0.0209$         & 
                                $0.4963\pm0.0402$         & $0.5361\pm0.0343$         & $0.1248\pm0.0117$         & 
                                $0.3066\pm0.0258$         & $0.2490\pm0.0274$        \\ \bottomrule
\end{tabular}%
}
\caption{Performance comparison across different models on Adamson and Norman datasets, evaluated by RMSE, Direction Accuracy Coefficient (DAC), and Pearson Correlation Coefficient (PCC) metrics.}
\label{tab:performance_comparison}
\end{table*}

\subsection{Genetic Perturbation Prediction}
During training process, \textbf{GRAPE} initializes gene embeddings $Z_{m}$ and constructs gene regulation network $\mathbb{G}$ at first. Given a gene perturbation condition $c$, which biologically corresponds to the knockout of a specific gene , we incorporate this perturbation information into the original input in multi-head HGAT to simulate its effects. Specifically, this involves masking the representations of the corresponding knockout genes in $Z_{\tt{fus}}$, ultimately resulting in the final output $P_{c}= \{ p_{c}^{1},p_{c}^{1},\dots,p_{c}^{N}  \} \in R^{N \times d}$ of Eq.~(\ref{eq:HGAT}).\\
\indent Considering that each gene has its own unique expression pattern \cite{GEARS}, we apply an independent linear transformation to each gene to obtain the final output under perturbation $c$:
\begin{equation}
    \hat{g}_{c}^{i}=W_{i}\odot p_{c}^{i}+b_{i} \in R
    \label{eq:gene_specific}
\end{equation}
where $\odot$ denotes Hadamard Product. Ultimately, the prediction of gene expression can be obtained by:
\begin{equation}
    \tilde{g}_{c}=\hat{g}_{c}+ctrl
    \label{eq:decoder}
\end{equation}
where $ctrl$ represents the expectation of gene expression levels from the control group cells (non-knockout).\\
\indent  To optimize the \textbf{GRAPE}’s performance, we compute the mean squared error (MSE) between the predicted and true gene expression values under condition $c$: 
\begin{equation}
    \mathcal{L}_{mse}=\frac{1}{N}\sum_{i=1}^{N}( g_{c}^{i} - \tilde{g}_{c}^{i} )^2
    \label{eq:MSE}
\end{equation}
where $g_{c}\in R^{N \times 1}$ represents the observed gene expression value sampled under perturbation condition $c$.
\indent Moreover, to control the direction of gene expression changes, we use the Huber loss function, which combines MSE's sensitivity to small errors with MAE's robustness to large errors, reducing the impact of outliers and ensuring more stable training in noisy data.
\begin{equation}
    \mathcal{L}_{dir}=\frac{1}{N}\sum_{i=1}^{N}Huber( tanh(g_{i}-ctrl_{i}),tanh(\tilde{g}_{i}-ctrl_{i}) )
    \label{eq:dir}
\end{equation}
\begin{equation}
    Huber(a,b)=\begin{cases}
                \frac{1}{2}(a-b)^2, & \text{if } |a-b| \leq \delta \\
                \delta (|a-b|-\frac{1}{2} \delta), & \text{otherwise}
                \end{cases}
    \label{eq:Huber}
\end{equation}
\indent The reconstruction objective of the model for gene expression values is represented as:
\begin{equation}
    \mathcal{L}_{recon}=\mathcal{L}_{mse}+\mathcal{L}_{dir}
\end{equation}

\subsection{Training Steps}
During model training, we optimize the above objectives simultaneously, and the final training objective is defined as:
\begin{equation}
    \mathcal{L}=\mathcal{L}_{recon}+\mathcal{L}_{reg}+\mathcal{L}_{align}
    \label{eq:final objective}
\end{equation}
here, $\mathcal{L}_{align}$is specifically optimized during the pre-training phase before the main model training.

\subsection{Analysis of Time Complexity}
To simplify the analysis, we assume the feature dimension has no impact and treat computations as $O(1)$. The computational cost of the model consists of four main components: multi-head attention $O(H_{1} \cdot N^2)$, $K$-hop similarity (\(O(K \cdot m )\)), channel attention $O(N^2)$, and feature extractions from multi-head HGAT $O(L \cdot H_{2} \cdot (|\mathbb{E}|+N) )$. Here, $N$ denotes the number of genes, $H_{1}$ represents head number in Eq.~(\ref{eq:K-head similarity}), $m$ stands for the number of edges in the result of Eq.~(\ref{eq:multi-view graph construction}), with $L$ and $H_{2}$ referring to the layer depth and the number of attention heads in Eq.~(\ref{eq:HGAT}), respectively. 

\begin{figure*}[htbp]
    \centering
    \includegraphics[width=\textwidth]{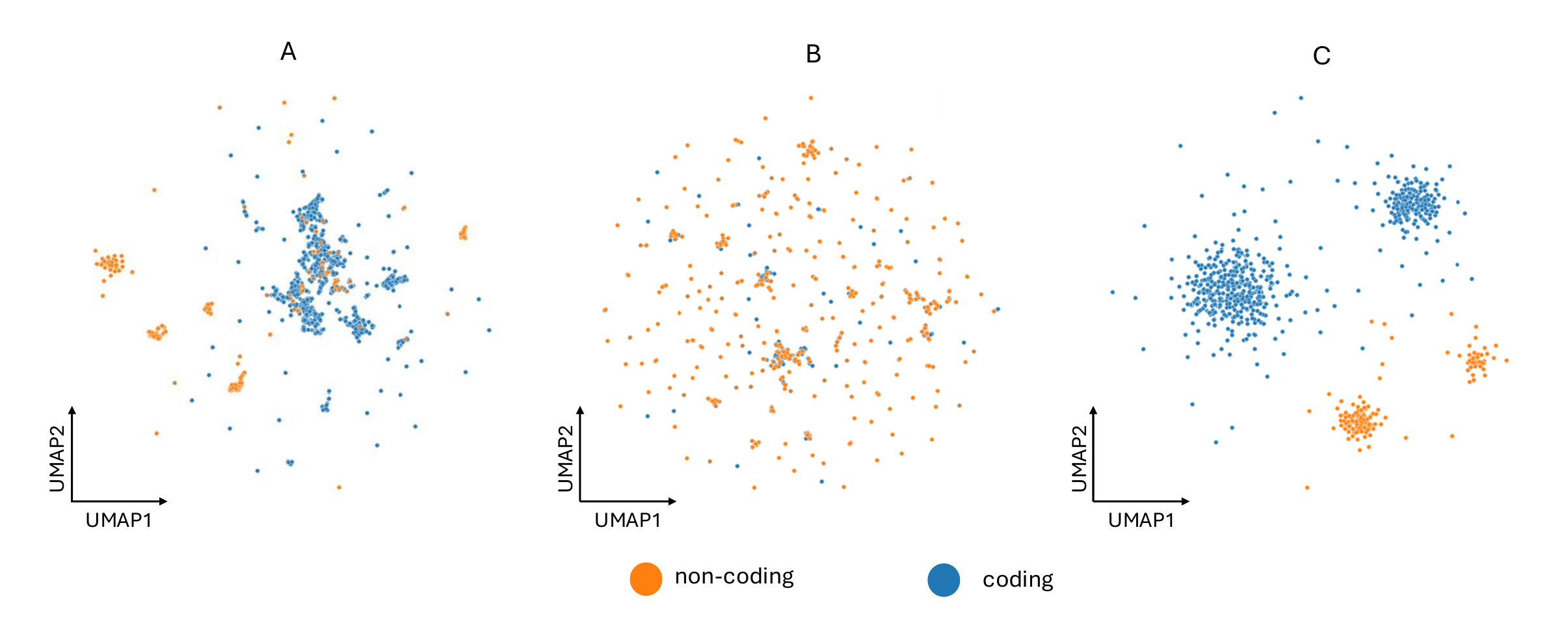}
    \caption{The UMAP visualization of gene representations. Subfigures A and B show the gene text modality representations and gene sequence modality representations extracted by the pre-trained LLM and DNA sequence model, respectively. Subfigure C illustrates the representations extracted by \textbf{GRAPE} after training.}
    \label{fig:UMAP}
\end{figure*}

\begin{figure}[htbp]
    \centering
    \includegraphics[width=0.48\textwidth]{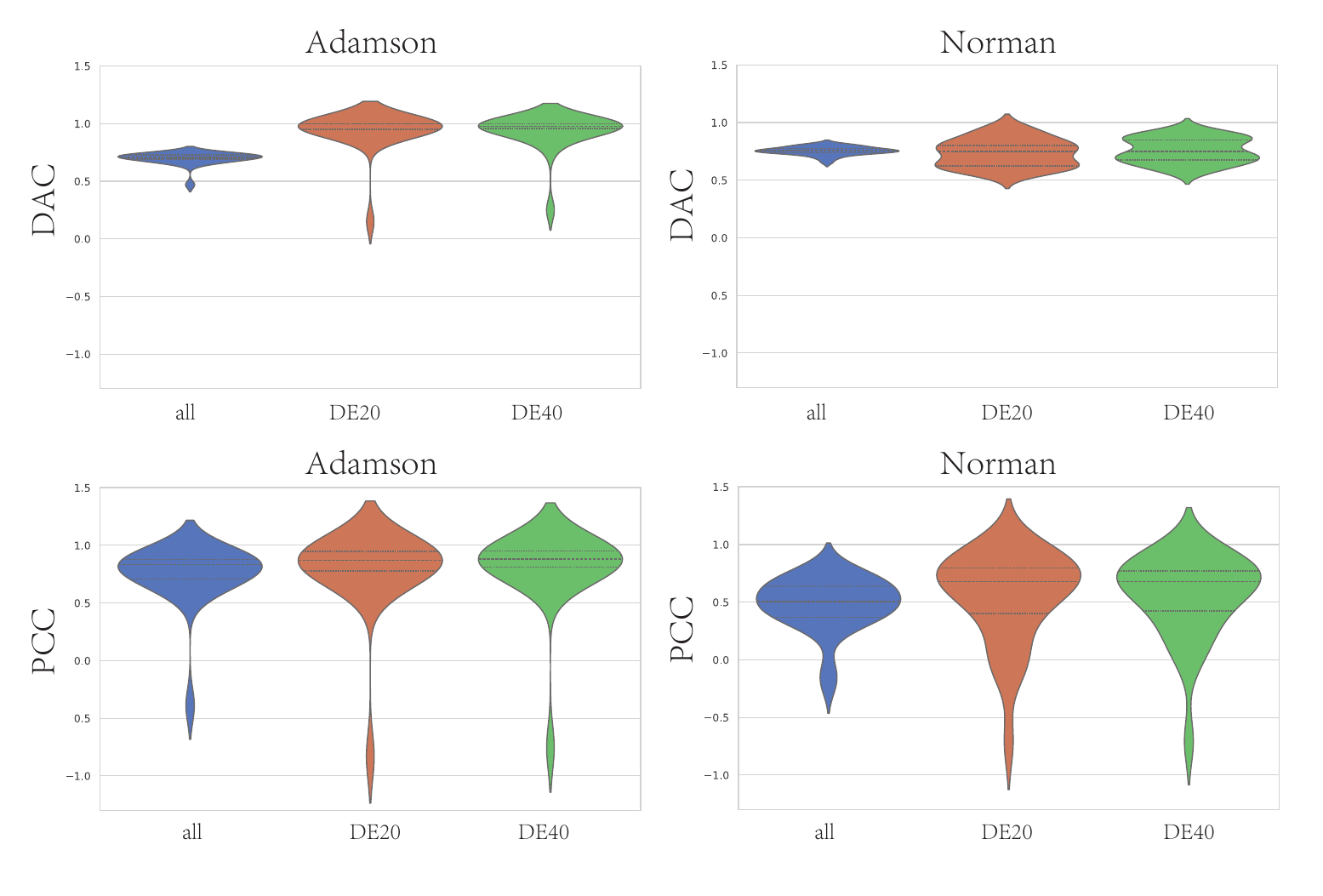}
    \caption{The picture shows the distribution of predicted values across samples.}
    \label{fig:violin}
\end{figure}

\section{Results and Discussion}
To demonstrate the effectiveness of our method, we conduct extensive experiments on public datasets. Adamson \cite{adamson} contains data from 87 types of single-gene perturbations, with a total of $68,603$ single-cell RNA sequencing results. Norman \cite{norman} includes results from both single-gene and double-gene perturbations. We only select experimental data for single-gene perturbations, with over 100 types of single-gene perturbations in total. In each dataset, hundreds of single-cell RNA sequencing data can be obtained for each type of genetic perturbation. 

\subsection{Experiment Settings}
Gene text descriptions, sequences, and biotypes can all be accessed through the API provided by Ensemble (\href{http://ensemblgenomes.org/}{Click here to visit}). In the training process of \textbf{GRAPE}, we randomly select 70\% of gene perturbation conditions for the training set and use the remaining for testing, then select single-cell data according to these conditions. The number of heads in both the multi-head similarity module (Eq.~\ref{eq:K-head similarity}) and the multi-head HGAT module (Eq.~\ref{eq:HGAT}) is set to 4. The model is trained using the AdamW \cite{AdamW} optimizer with a learning rate of 0.001 and a batch size of 64. For evaluation, we adopt multiple metrics, including Direction Accuracy Coefficient (DAC), Root Mean Square Error (RMSE), and Pearson Correlation Coefficient delta (PCC\_delta), to assess the model's prediction accuracy and biological relevance (refer to the supplementary materials for further details). All our method and its competitors are conducted using four Nvidia A100 GPUs.

\subsection{\textbf{GRAPE} outperform existing methods}
Table \ref{tab:performance_comparison} demonstrates that \textbf{GRAPE} significantly outperforms existing methods across all evaluated metrics. While Root Mean Square Error (RMSE) measures the overall prediction error, DAC and PCC\_delta provide deeper biological insights. DAC evaluates the consistency between predicted and true directions of gene expression changes, which is crucial for understanding gene regulatory effects and ensuring accurate biological interpretations, as incorrect directional predictions can lead to misleading conclusions. PCC\_delta, on the other hand, quantifies the consistency in the change of gene expression values between the predicted and true values under perturbation conditions. Together, these metrics offer a comprehensive assessment of model accuracy and relevance in biological contexts.\\
\indent \textbf{GRAPE} showcases superior performance compared to GEARS \cite{GEARS}, primarily because it goes beyond GEARS' simplified gene interaction network by accounting for the diverse relationships across different biotype genes. \textbf{GRAPE} actively uncovers latent connections between genes, providing a deeper understanding of gene regulatory dynamics. In contrast, generative models \cite{graphVCI,sams-vae} perform poorly due to their inability to effectively capture the semantic meaning of gene perturbation conditions. This limitation hinders their ability to model biologically relevant features, resulting in less accurate predictions. Moreover, \textbf{GRAPE} leverages multi-modal information to initialize gene representations, providing a more comprehensive and biologically meaningful characterization of genes.\\
\indent To further evaluate the model's performance, we visualize the distribution of DAC and PCC values across samples (Fig.\ref{fig:violin}). These plots provide a detailed view of the prediction accuracy at the sample level, highlighting the variability and consistency of the model’s predictions. By focusing on these key metrics, the violin plots offer an intuitive representation of how well the model captures the directionality and correlation of gene expression changes under different perturbations.

\begin{figure*}[htbp]
    \centering
    \includegraphics[width=1.02\textwidth]{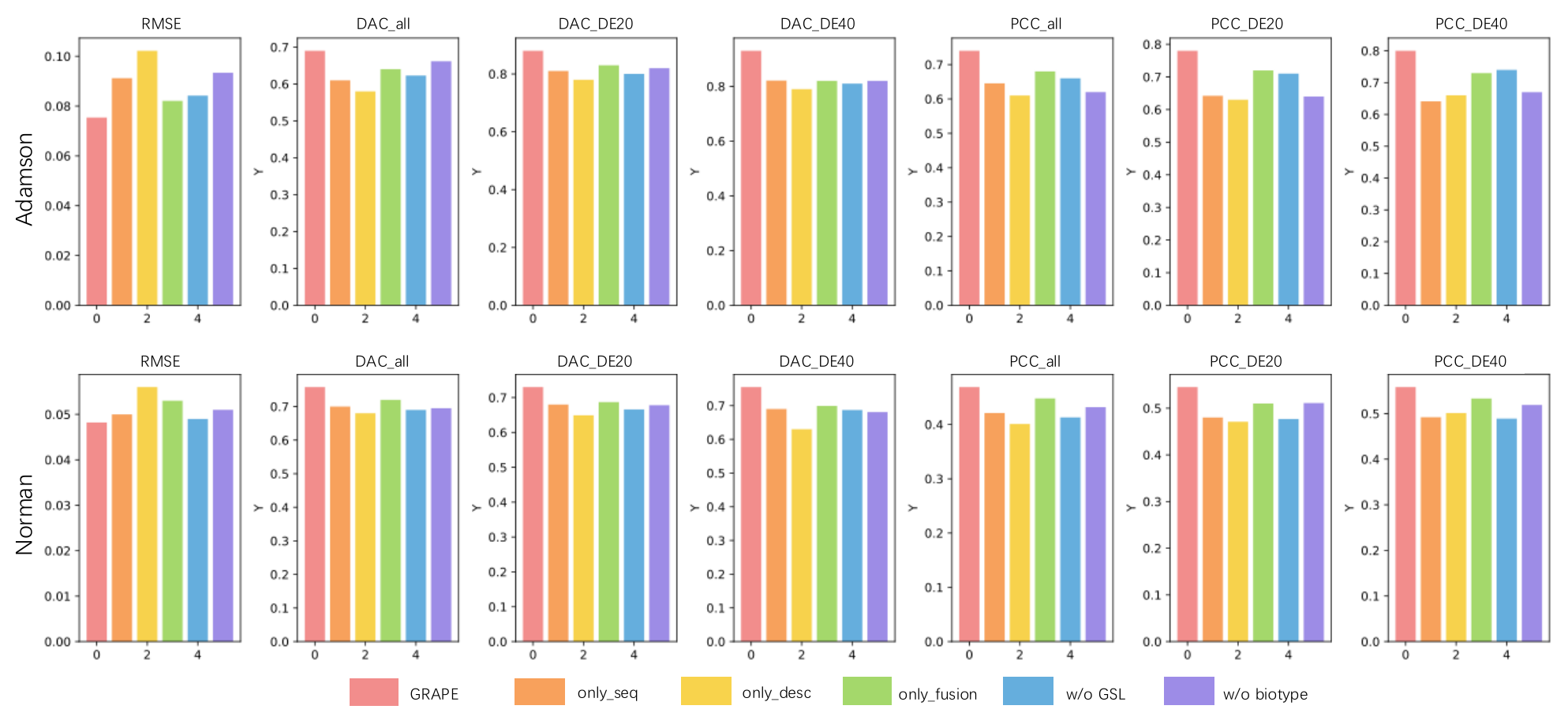}
    \caption{Ablation Study Results.}
    \label{fig:BarPolt}
\end{figure*}

\subsection{\textbf{GRAPE} Enhances Gene Representation with Biotype-Aware Heterogeneous Graphs}
To evaluate \textbf{GRAPE} leveraging biotype knowledge, we first visualize gene features extracted from pre-trained LLM \cite{bert} and DNA sequence model \cite{hyenaDNA}, focusing on two modalities: gene textual descriptions (Fig.\ref{fig:UMAP}.A) and gene sequences (Fig.\ref{fig:UMAP}.B). For this, we employe Uniform Manifold Approximation and Projection (UMAP) to visualize the high-dimensional gene features in a lower-dimensional space. The visualizations of Fig.\ref{fig:UMAP}.A and Fig.\ref{fig:UMAP}.B reveal that the clustering of gene representations did not align with the biotype labels. Despite the inherent complexity and richness of these features, the unsupervised clustering approach failed to exhibit any clear, biotype-specific grouping. This suggests that the pre-trained models, although capable of extracting meaningful features, does not sufficiently capture the underlying biological relevance of gene biotypes on its own. \\
\indent In contrast, after training the \textbf{GRAPE}, which incorporates biotype-aware heterogeneous graphs into the representation learning process (Eq.~(\ref{eq:HGAT})), the extracted gene features exhibited a distinct biotype-based clustering pattern (Fig.\ref{fig:UMAP}.C). The representations of genes that belong to the same biotype formed tight and well-separated clusters, indicating that the model successfully learned to embed gene biotype information into the feature space. This outcome highlights the efficacy of \textbf{GRAPE} in improving gene representation by dynamically adjusting the gene regulatory network and then extracting features through HGAT that incorporates biotype-aware information.\\
\indent Constructing a heterogeneous graph network effectively represents the complex relationships between genes and simulates interactions between different biotype genes, thereby enhancing gene representation through HGAT. On the other hand, when biotype is directly concatenated as a feature to the gene representation, the model may fail to fully explore the underlying connections between genes based on their biotype. This approach only provides a singular biotype feature, lacking interactions with other gene attributes and fails to leverage the diversity of nodes and edges in the GRN. As a result, the model may not capture the biological significance of genes in their multidimensional relationships. Additionally, directly incorporating the biotype label as a feature could lead to over-reliance on the biotype information, potentially neglecting other crucial biological features, which could in turn reduce the model's generalization ability.

\subsection{Ablation Study}
To further evaluate the effectiveness of \textbf{GRAPE}, we compare it with the following methods through an ablation study. 1)\textbf{only\_seq}: Excludes the textual description modality features, using only the DNA sequence features in the model. 2)\textbf{only\_desc}: Excludes the DNA sequence modality features, using only the textual description features in the model. 3)\textbf{only\_fusion}: Uses the fusion modality, combining both desc and seq features into a unified representation, excluding the individual modalities. 4)\textbf{w/o~GSL}: Excludes the K-hop similarity graph structure learning process. 5)\textbf{w/o~biotype}: Ignores biotype information, converting the GRN into a homogeneous graph, and uses a homogeneous GNN instead of HGNN. The results are shown in Figure.\ref{fig:BarPolt}.\\
\indent The experimental results show that \textbf{GRAPE} outperforms methods relying on individual modalities. This indicates that integrating multiple modalities allows the model to capture a broader range of information, leveraging complementary insights from each modality for improved performance and more robust feature extraction. Additionally, the performance of \textbf{no\_GSL} is less than ideal, highlighting the significant role that graph structure learning plays in capturing the hidden relationships between genes. Without graph structure learning, the model fails to fully leverage the intricate connections and dependencies inherent in the gene interactions, which are essential for accurate representation and prediction.\\
\indent More importantly, the results clearly demonstrate the effectiveness of introducing biotype information. By leveraging HGNN, the model is able to capture and simulate the complex, multi-dimensional regulatory relationships between genes. This approach not only allows for a more nuanced understanding of gene interactions across different biological categories, but also enhances the model's ability to predict gene behavior with greater accuracy.

\section{Conclusion}
In this work, we present \textbf{GRAPE}, a heterogeneous graph neural network (HGNN) that integrates gene biotype information for predicting genetic perturbations. By leveraging pre-trained large language model (LLM) and DNA sequence model to extract features from gene textual descriptions and DNA sequences, we initialize gene representations in a way that captures both sequence-based and textual semantic information. Furthermore, for the first time, we incorporate gene biotype data to simulate the distinct roles of different biotypes in regulating cellular processes. Through graph structure learning (GSL), \textbf{GRAPE} dynamically refines the gene regulatory network (GRN), effectively capturing implicit gene relationships. Our method outperforms existing approaches on publicly available datasets, achieving state-of-the-art performance in predicting genetic perturbations. If sufficient data is available, it can enable a zero-shot model on top of \textbf{GRAPE}, automatically constructing gene regulatory networks (GRN) and inferring gene interactions to predict genetic perturbation results without human annotations.


\bibliographystyle{named}
\bibliography{ijcai25}

\end{document}